\documentclass[final,5p,times,twocolumn]{elsarticle}
\usepackage{amssymb}
\usepackage{graphicx}
\usepackage{rotating}
\usepackage{datetime}
\journal{Physics Letters B}
\begin{document}
\begin{frontmatter}
\title{High-resolution measurement of the time-modulated orbital electron capture and of the $\beta^+$ decay of hydrogen-like $^{142}$Pm$^{60+}$ ions\\
{\small(Revised version: {\settimeformat{ampmtime}\today, \currenttime)}}}
\date{today}
\author[TUM]{P.~Kienle\dag}%
\author[GSI]{F.~Bosch}%
\author[VIE]{P.~B\"uhler}%
\author[TUM]{T.~Faestermann}%
\author[GSI,MPI,UniH]{Yu.A.~Litvinov}%
\author[GSI,UniF]{M.~S.~Sanjari}%
\author[MPI,UniH]{D.~B.~Shubina}%
\author[GSI,MPI]{N.~Winckler}%
\author[GSI,MPI,UniH]{D.~Atanasov}%
\author[GSI,JLU]{H.~Geissel}%
\author[GSI]{V.~Ivanova}%
\author[MPI,IMP]{X.~L.~Yan}%
\author[GSI,FRA]{D.~Boutin}%
\author[GSI,EMMI,IAP]{C.~Brandau}%
\author[GSI,JLU]{I.~Dillmann}%
\author[GSI]{Ch.~Dimopoulou}%
\author[GSI,EMMI]{R.~Hess}%
\author[GSI,IAP]{P.-M.~Hillebrand}%
\author[Nii]{T.~Izumikawa}%
\author[GSI,JLU]{R.~Kn\"obel}%
\author[GSI,CERN]{J.~Kurcewicz}%
\author[JLU]{N.~Kuzminchuk}%
\author[GSI]{M.~Lestinsky}%
\author[GSI]{S.~A.~Litvinov}%
\author[IMP]{X.~W.~Ma}%
\author[TUM]{L.~Maier}%
\author[GSI,ITA]{M.~Mazzocco}%
\author[GSI]{I.~Mukha}%
\author[GSI]{C.~Nociforo}%
\author[GSI]{F.~Nolden}%
\author[GSI,JLU]{Ch.~Scheidenberger}%
\author[GSI]{U.~Spillmann}%
\author[GSI]{M.~Steck}%
\author[GSI,UniJ,HIJ]{Th.~St\"ohlker}%
\author[GSI,JLU,Bei]{B.~H.~Sun}%
\author[UniS]{F.~Suzaki}%
\author[UniS]{T.~Suzuki}%
\author[SPB]{S.~Yu.~Torilov}%
\author[GSI,Paris]{M.~Trassinelli}%
\author[GSI,IMP]{X.~L.~Tu}%
\author[MPI,IMP]{M.~Wang}%
\author[GSI]{H.~Weick}%
\author[GSI]{D.~F.~A.~Winters}%
\author[GSI,UniH]{N.~Winters}%
\author[UniE]{P.~J.~Woods}%
\author[UniS]{T.~Yamaguchi}%
\author[Bei]{G.~L.~Zhang}%
\author{\newline\textbf{(the~Two-Body-Weak-Decays~Collaboration)}}%

\address[TUM]{Technische Universit{\"a}t M{\"u}nchen, 85748 Garching, Germany}%
\address[GSI]{GSI Helmholtzzentrum f\"ur Schwerionenforschung, 64291 Darmstadt, Germany}%
\address[VIE]{Stefan Meyer Institut f{\"u}r subatomare Physik, 1090 Vienna, Austria}%
\address[MPI]{Max-Planck-Institut f\"ur Kernphysik, 69117 Heidelberg, Germany}%
\address[UniH]{Ruprecht-Karls Universit{\"a}t Heidelberg, 69120 Heidelberg, Germany}%
\address[UniF]{J.~W.-Goethe Universit{\"a}t, 60438 Frankfurt, Germany}%
\address[JLU]{II. Physikalisches Institut, Justus-Liebig Universit{\"a}t, 35392 Gie{\ss}en, Germany}%
\address[IMP]{Institute of Modern Physics, Chinese Academy of Sciences, Lanzhou 730000, China}%
\address[FRA]{Service de Physique Nucl\'eaire, CEA-Saclay, F-91191 Gif-Sur-Yvette, Cedex, France}%
\address[EMMI]{ExtreMe Matter Institute EMMI, 64291 Darmstadt, Germany}%
\address[IAP]{Institut f{\"u}r Atom- und Molek{\"u}lphysik, Justus-Liebig Universit{\"a}t, 35392 Gie{\ss}en, Germany}%
\address[Nii]{Radio Isotope Center, Niigata University, Niigata 951-8510, Japan}%
\address[CERN]{CERN, 1211 Geneva 23, Switzerland}%
\address[ITA]{Dipartimento di Fisica, INFN, I35131 Padova, Italy}%
\address[UniJ]{Friedrich-Schiller-Universit{\"a}t Jena, 07737 Jena, Germany}%
\address[HIJ]{Helmholtz-Institut Jena, 07743 Jena, Germany}%
\address[Bei]{School of Physics \& Nucl. Energy Engineering, Beihang Univ., 100191 Beijing, China}%
\address[UniS]{Graduate School of Science \& Engineering, Saitama Univ., Saitama 338-8570, Japan}%
\address[SPB]{St. Petersburg State University, 198504 St. Petersburg, Russia}%
\address[Paris]{INSP, CNRS and Universit{\'e} Pierre et Marie Curie, UMR7588, 75005 Paris, France}%
\address[UniE]{School of Physics \& Astronomy, The University of Edinburgh, Edinburgh EH9 3JZ, UK}%
%

\begin{abstract}
The periodic time modulations, found recently in the two-body orbital electron-capture (EC) decay of both, 
hydrogen-like $^{140}$Pr$^{58+}$ and $^{142}$Pm$^{60+}$ ions, 
with periods near to 7~s and amplitudes of about 20\%, 
were re-investigated for the case of $^{142}$Pm$^{60+}$ by using a 245~MHz resonator cavity with a much improved sensitivity and time resolution. 
We observed that the exponential EC decay is modulated with a period $T = 7.11(11)$~s, in accordance with a modulation period $T = 7.12(11)$~s 
as obtained from simultaneous observations with a capacitive pick-up, employed also in the previous experiments. 
The modulation amplitudes amount to $a_R = 0.107(24)$ and $a_P = 0.134(27)$ for the 245~MHz resonator and the capacitive pick-up, respectively. 
These new results corroborate for both detectors {\it exactly} our previous findings of modulation periods near to 7~s, 
though with {\it distinctly smaller} amplitudes. 
Also the three-body $\beta^+$ decays have been analyzed. For a supposed modulation period near to 7~s we found an amplitude $a = 0.027(27)$, 
compatible with $a = 0$ and in agreement with the preliminary result $a = 0.030(30)$ of our previous experiment. 
These observations 
could point at weak interaction as origin of the observed 7~s-modulation of the EC decay. 
Furthermore, the data suggest that interference terms occur in the two-body EC decay, although the neutrinos are not directly observed. 
\end{abstract}%

\end{frontmatter}%

\section{Introduction}
Recently we reported on the measurement of two-body orbital electron capture (EC) decays 
of hydrogen-like (H-like) $^{140}_{~59}$Pr$^{58+}$ and $^{142}_{~61}$Pm$^{60+}$ ions~\cite{Osc-PLB}. 
We observed for both ion species, coasting at a Lorentz factor $\gamma = 1.43$ in the ESR ion storage ring, modulations  
in the time dependence of the EC decays with periods near to 7~s and with modulation amplitudes of about 20\%. 
Both, the $^{140}_{~59}$Pr and the $^{142}_{~61}$Pm nuclei undergo $\beta^+$ or EC decay with pure Gamow-Teller transitions 
from $I = 1^+$ states to $I = 0^+$ nuclear ground states of stable $^{140}_{~58}$Ce and $^{142}_{~60}$Nd nuclei, respectively, 
with a branching of nearly 100\%.  
Using time-resolved Schottky Mass Spectrometry (SMS) of single cooled ions~\cite{Li-NPA04, Ge-NPA04, Li-NPA05, FGM}, 
the EC decay of an ion is unambiguously identified by a sudden rise $\Delta f$ 
of its revolution frequency $f$ in the order $\Delta f/f \approx 10^{-6}$, 
corresponding to the $Q_{EC}$ value of the EC decay, as the charge state of the ion does {\it not} change. 
In the previous experiment the recoiling daughter ions appeared delayed with respect to their generation, 
because they could be registered by SMS only after the completion of the electron cooling~\cite{ecool}. 
For the capacitive pick-up detector~\cite{Schaaf} used in these first experiments, the signal-to-noise ratio was too poor 
for observing the cooling process in detail. 
Thus we could determine only the appearance time of the cooled daughter ion, 
delayed by a cooling time that could only be roughly estimated at that time~\cite{Osc-PLB}.

A further noticeable result of the previous experiment was the observation, 
that the time-averaged decay probability of the modulated EC decays 
was fully consistent with the decay constant of neutral atoms 
after taking the atomic charge state into account~\cite{Li-PRL07,PK-PRC08, Patyk08, Patyk10, Wi-PLB,Atanasov}. 
This excluded the influence of possible oscillatory transitions with a period of 7~s between 
the $F=1/2$ hyperfine ground state of the H-like ions and their $F= 3/2$ excited state 
which is stable with respect to allowed weak decay~\cite{Wi-arxiv}. 

Also the simultaneously observed $\beta^+$ decays of H-like $^{142}$Pm$^{60+}$ ions were analyzed in this experiment. 
Those decays could be safely identified, however, only for a sub-set of observation cycles where at most two parent ions were stored in the ESR. 
According to preliminary data from this sub-set $(\approx1800~\beta^+$ decays), the three-body $\beta^+$ decay exhibits, 
for a supposed modulation period around 7~s, an amplitude of $a = 0.03(3)$, compatible with $a = 0$~\cite{PK-NPA09, PK-PPNP10}. 
These first observations made it indispensable to develop a new pick-up system 
with an improved signal-to-noise ratio and time resolution for scrutinizing our experimental results. 
Further reasons for improved measurements are the so far very controversial arguments on the origin of the modulation observed in the EC decay. 

In Ref.~\cite{Osc-PLB} it was tentatively proposed to interpret the modulations 
as due to the properties of the electron neutrino that is generated in the 
EC decay as a coherent superposition of  mass eigenstates. 
However, such a suggested connection of the modulations with quantum interference 
of neutrino mass eigenstates (see also Refs.~\cite{Lipkin-1, Lipkin-2, Lipkin-3}) 
has been vigorously criticized (see, e.  g., Refs.~\cite{Giunti, Kienert08, Merle10, Cohen}). 
Also several alternative physical explanations have been proposed (see, e. g., Refs.~\cite{alt1, alt2, alt3, alt4}). 


\section{Schottky-noise detection by means of a 245~MHz resonator}
Here we report results of the re-investigation of the EC- and $\beta^+$-decay of H-like $^{142}$Pm$^{60+}$ ions. 
For this purpose we installed in the storage ring 
in addition to the capacitive pick-up~\cite{Schaaf} a newly designed Schottky-noise frequency detector~\cite{Nolden}. 
This device, a 245~MHz pillbox-like resonator cavity, exhibits a signal-to-noise ratio improved by about two orders of magnitude 
with respect to the capacitive pick-up and reveals in an unprecedented manner hitherto hidden details of the EC decay of {\it a single ion}. 
It provides not only the true decay time, but also the time-and frequency-resolved kinematics and the entire cooling process of the recoiling daughter ion, 
just from the moment of its generation.
%
\begin{figure}[t!]
\begin{center}
\centering\includegraphics[angle=-0,width=0.45\textwidth]{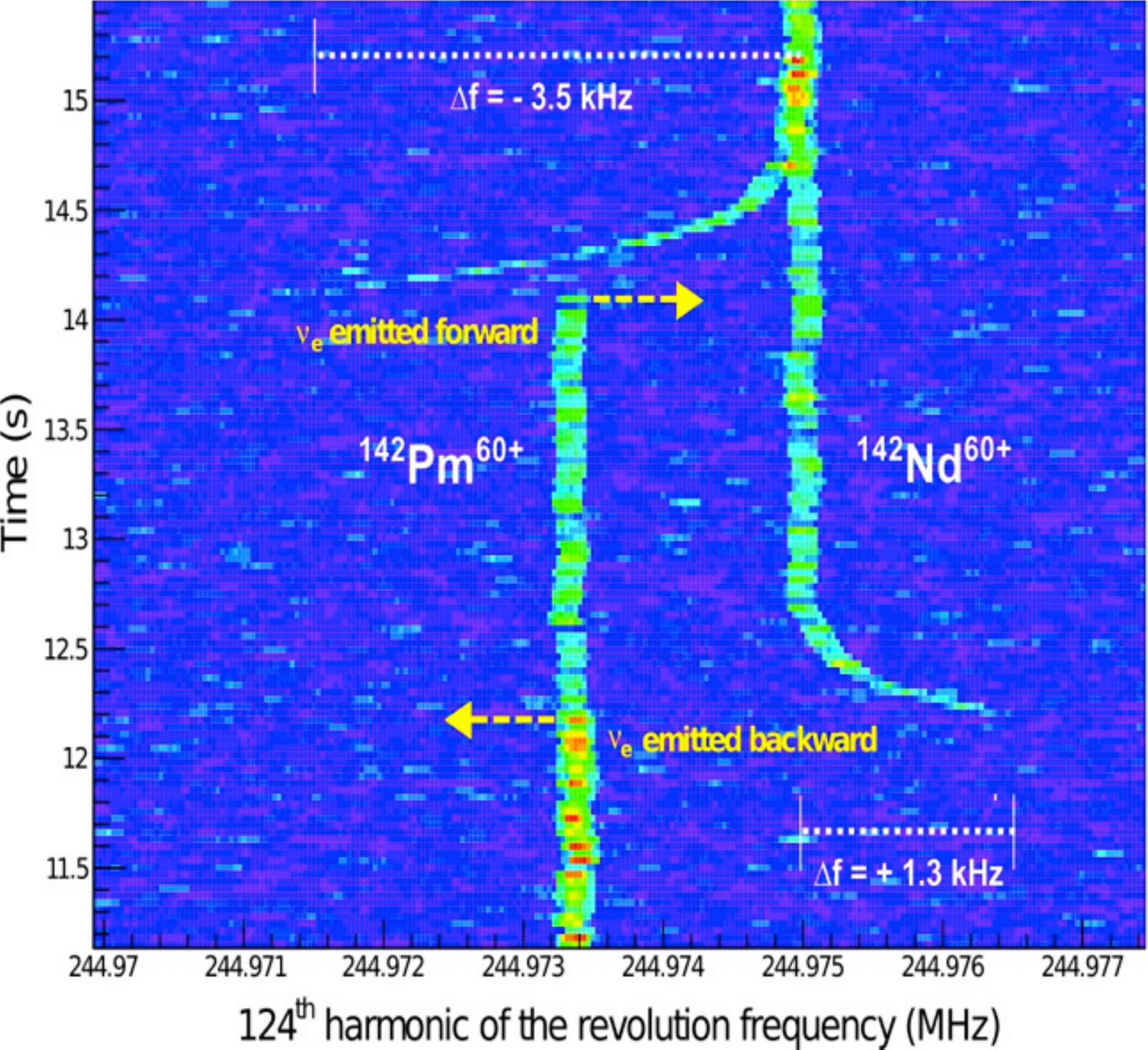}
\caption{
(Color online) Traces of two cooled $^{142}$Pm$^{60+}$ parent ions, recorded at the 124$^{\rm th}$ harmonic of the revolution frequency 
by the 245~MHz resonator vs. the time after injection, 
with time- and frequency-binning of 32~ms and 31.25~Hz, respectively. 
Both parent H-like $^{142}$Pm$^{60+}$ ions decay by EC to bare $^{142}$Nd$^{60+}$ daughter ions
accompanied by the emission of a ``monoenergetic'' electron neutrino $\nu_e$. 
The yellow arrows indicate the true decay times, as unambiguously 
identified by a decrease of the intensity of the trace of the parent ions 
and the {\it simultaneous} onset of the trace of the recoiling daughter ion. 
The latter starts at a revolution frequency shifted by $\Delta f$ with respect 
to the frequency after completion of electron cooling, which reflects the 
projection of the recoil velocity onto the beam direction axis immediately after the decay.
}%
\label{spectrum3d}%
\end{center}
\end{figure}
%

Figure~\ref{spectrum3d} shows a part of the time vs. revolution-frequency spectrum of two injected parent ions 
with time- and frequency-binning of 32~ms and 31.25~Hz, respectively. 
Both parent ions decay by orbital EC where a ``monochromatic'' electron neutrino $\nu_e$ will be emitted.
The spectrum was recorded by the 245~MHz resonator at the 124$^{\rm th}$ 
harmonic of the 1.98~MHz revolution frequency of the parent ions. 
Those EC decays are identified by a frequency rise of 1560~Hz for the cooled daughter ions 
with respect to the frequency of the cooled parent ions, corresponding to a $Q_{EC}$ value of 4830~keV. 

The neutrino transfers an energy of about 90~eV (centre of mass system) to the daughter ion, 
which recoils with a velocity ${\vec v}_R/c = - {\vec p}_{\nu}c/M_dc^2$, 
where ${\vec p}_{\nu}$ is the momentum of the monoenergetic electron neutrino and $M_d$ is the mass of the daughter ion. 
Depending on the emission angle of the neutrino, the corresponding velocity ${\vec v}_R$ of the recoiling daughter ion 
leads to a shift $\Delta{f}$ of its revolution frequency  to values between $f \pm\Delta{f}$, 
where $f$ denotes its revolution frequency after the completion of cooling. 
The difference $\Delta{f}$ between the revolution frequency of the recoiling daughter ion immediately at the decay 
and after completion of cooling reflects {\it the projection of its initial velocity onto the beam axis}.  
The maximum {\it observed} frequency shift $\Delta{f} (max) = \pm 3.91$~kHz, which we assign to the strict forward 
or backward emission of the recoiling daughter ion relative to the beam direction (vice versa of the neutrino), 
provides the modulus of the recoil velocity. 
Hence, for each observed $\Delta{f}$ the polar emission angle of the daughter ion (and of the neutrino) can be derived.  
If the beam is (longitudinally) not polarized, a {\it uniform} distribution of $\Delta{f}$ in between $-3.91~{\rm kHz}   \le \Delta f \le + 3.91~{\rm  kHz}$ 
has to be expected for an {\it isotropic} emission of the electron neutrino in the centre of mass system.  
All these features have been observed in the present experiment by observing many thousands of EC decays. 

As shown in Fig.~\ref{spectrum3d}, the complete cooling processes of the two recoiling $^{142}$Nd$^{60+}$ ions, 
following the emission of the neutrino into the backward or forward hemisphere, are revealed by the 245~MHz resonator.  
In particular, the true decay time, as indicated by the onset of the cooling trace of the daughter ion 
and the {\it simultaneous} intensity decrease of the trace of the cooled parent ions, 
can be determined with an accuracy of approximately 32~ms.  

\section{Experimental facts and data analysis}
\label{s:s3}
As in the previous experiment~\cite{Osc-PLB}, the H-like $^{142}$Pm$^{60+}$ ions 
were produced by fragmentation of 600~$A\cdot$MeV $^{152}$Sm ions in a 2.5~g/cm$^2$ Be-target. 
The primary $^{152}$Sm beam was provided by the SIS18 heavy-ion synchrotron. 
The pure beams of $^{142}$Pm$^{60+}$ fragments were separated in-flight by means of the fragment separator FRS and injected into the ESR storage ring.
Many thousands of injections of the ions were done during this experiment. 

The injection into the ring involves a kicker magnet which ``kicks'' the ions onto a closed orbit in the ESR.
If the already stored ions receive a kick, they will be removed from the ring.
Therefore, the synchronization and the duration of the kicker pulse is indispensable.
As a result, every time 
a 300~ns long bunch at 400~$A\cdot$MeV (Lorentz factor $\gamma = 1.43$) of $^{142}$Pm$^{60+}$ions was injected. 
The velocity spread of about 1\% from the fragmentation was reduced within the first 3.5~seconds by stochastic cooling~\cite{scool}
followed by electron cooling~\cite{ecool} with an electron current of 250~mA to a relative velocity spread $\Delta{v}/v \approx 5\cdot10^{-7}$. 
On average three H-like $^{142}$Pm$^{60+}$ ions were injected and monitored by both the capacitive 
pick up and the 245~MHz resonator for an optimized measuring period of 54~s with sampling times of 64~ms and 32~ms, respectively. 
The noise-power signals induced in the pick-up plates were amplified, digitized, Fourier transformed and stored for off line analysis. 
In the case of the 245~MHz resonator, the raw sampled data were stored, which enabled off-line Fourier analysis.

The ions circulating in the storage ring were removed in the previous experiment after each observation period 
by means of a mechanical shutter which was closed and opened for each new injection. 
Since the many thousands injections planned for the present experiment appeared too risky 
for the mechanical stability of components of the beam shutter in vacuum, 
we relied in the present experiment upon the proper operation of the injection kicker, 
which should simultaneously inject new ions and remove the remaining ions of the preceding injection. 
This requires an exact synchronization between the extraction of the ions from 
the synchrotron and the injection into the storage ring as well as a sufficient duration of the kicker pulse. 
In the present experiment a manual adjustment of these timings was required, which was unfortunately reset 
by the accelerator control software at each failure of the machine hardware, e. g., ESR magnets.
Missing or incorrect re-adjustments of the kicker timing could lead to measured spectra, in which  
parent ions from previous injections were still present.  
In this case the phase of a potential modulation, which is related to the time of injection of the parent ions, could be obscured.

Altogether 17460 injection cycles have been performed, where on average four parent ions have been injected. 
The whole set of 8665 recorded EC decays, irrespective of the kicker-timing conditions, does {\it not} show any significant modulations.
One long series of 7125 {\it consecutive} injections with 3594 EC decays could be established, 
where no failures of the ESR hardware--relevant for the kicker timing--was documented in any of the experiment- or accelerator protocols.
%
This series has been used for the data analysis described below. 
Systematic time-differential scans of the region outside of the 7125 consecutive injections, 
where altogether eleven breaks of ESR magnets were registered, did {\it not} yield significant modulations.
However, neither in the analyzed data set of 7125 consecutive injections nor elsewhere in the data 
does the possibility exist to determine injection-by-injection whether the kicker-timing was set correctly 
and thereby make sure that all stored ions were removed after each cycle.   

All EC decay-spectra from both detectors were analyzed visually and computer assisted by independent groups (section~\ref{s41}). 
In addition the decay spectra of the 245~MHz resonator were investigated with respect to the time distribution of the $\beta^+$ decays (section~\ref{s42}). 

The EC decay rates displayed in Figs.~\ref{fig2} and \ref{fig3} comprise the number $dN(t)$ of EC decays 
per time interval $dt$, where the time $t$ is related to the time of injection, $t = 0$. 
Two models for describing the time distribution of these EC decays have been considered and fitted, namely a {\it strictly exponential distribution} (model $M_0$):
\begin{equation}
dN_d(t)/dt = \lambda_{EC} N_m(0) e^{-\lambda{t}}
\label{eq:eq1}
\end{equation}
 and a {\it modulated exponential function} (model $M_1$)  in which a  periodic time modulation of the EC decay-parameter $\lambda_{EC}$ is assumed:
\begin{equation}
\lambda_{EC}(t) = \lambda_{EC} [1 + a\cdot \cos (\omega{t} + \phi)]
\label{eq:eq2}
\end{equation}

Here $\lambda_{EC}$ is the EC decay constant, $a$ the time independent amplitude, $\omega = 2\pi/T$ the angular frequency, 
and $\phi$ the phase of a modulation with period $T$. $N_m(0)$ is the number of H-like parent ions at injection $(t = 0)$, $\lambda$  
the total decay parameter with $\lambda = \lambda_{\beta^+} + \lambda_{EC} + \lambda_{\rm loss}$, 
where $\lambda_{\rm loss}$ is the small loss constant of the coasting  ions in the ring. 
The latter is due to the interaction of the stored ions with the residual gas and the electrons of the cooler and 
amounts in the present experiment to $\lambda_{\rm loss} = 5(1) \cdot 10^{-4}$~s$^{-1}$. 
In the case of the 245~MHz resonator, the EC- and $\beta^+$- decay patterns were first simultaneously fitted by 
pure exponential functions with a {\it common} total decay constant $\lambda$. 
In the further analyses and fits of the individual decay curves, the total decay constant $\lambda$ was kept fixed at this value.

All the fits have been carried out also with floating $\lambda$. 
It was found that this had only a marginal effect on the fitted values of the other parameters as well as on the $\chi^2$-values. 
Since naturally the total decay constant $\lambda$ of both, the EC and $\beta^+$ decay of the parent ions has to be the same, 
we think that it is justified to determine the total decay constant with a fit to both decay curves and then keep this value fixed for all further steps. 
With this we eliminated an additional, unnecessary fit parameter without changing the principal results.

The fits were performed by means of $\chi^2$ minimization as well as by unbinned maximum likelihood estimations, which yielded consistent results. 
The variances of the modulation parameters of the EC decays and their correlations were studied in detail and scrutinized by extended simulations. 
In particular, no significant dependence of the modulation parameters on the time window of the analyses or the binning of the data was found.

\section{Results}
\subsection{Results for the EC decay of H-like $^{142}$Pm$^{60+}$ ions}
\label{s41}
\begin{figure}[t!]
\begin{center}
\centering\includegraphics[angle=-0,width=0.45\textwidth]{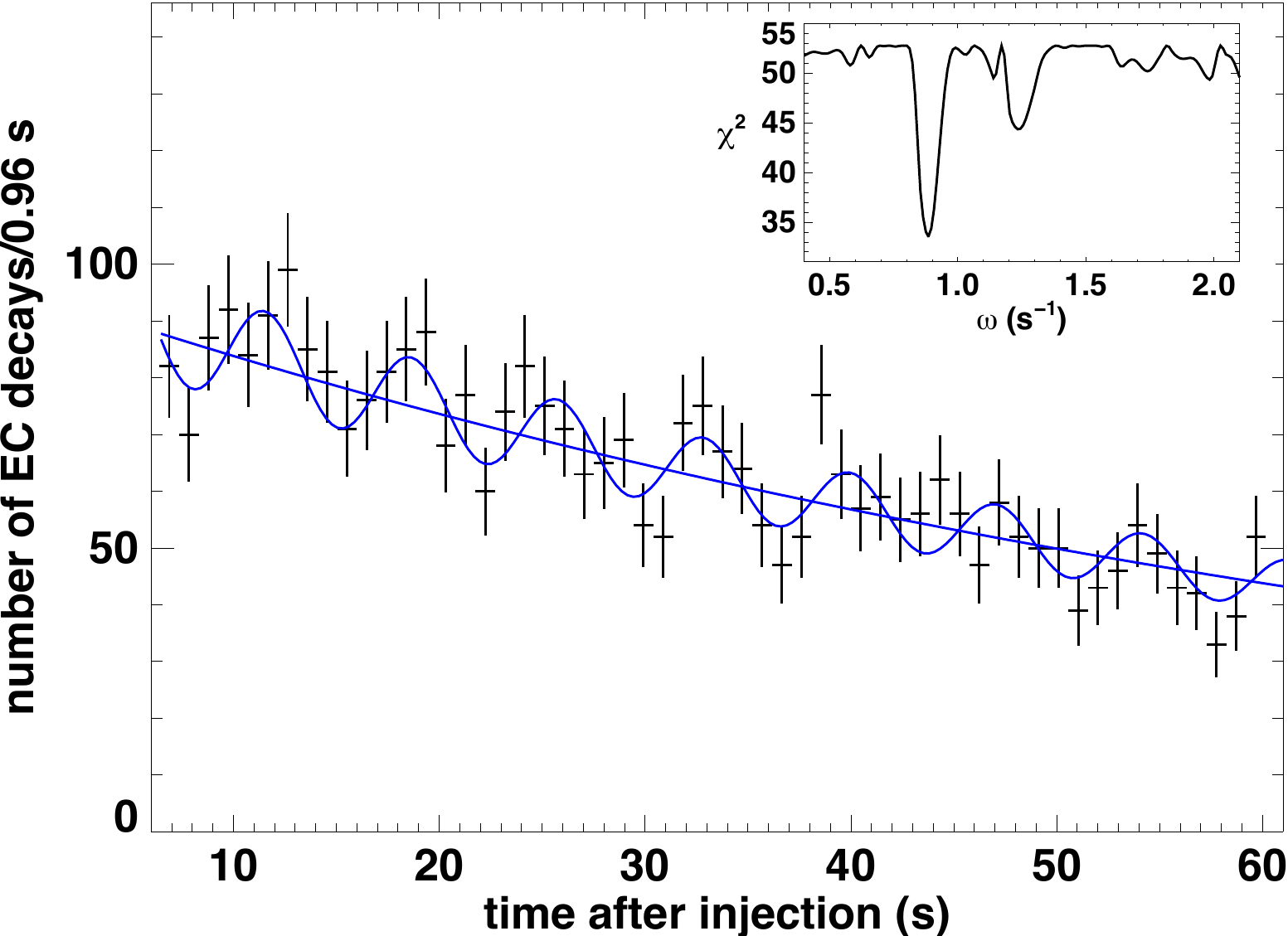}
\caption{
Number of EC decays per 0.96~s of H-like $^{142}$Pm$^{60+}$ ions, recorded by the 245~MHz resonator, 
vs. the time after injection of the ions into the storage ring ESR. 
Displayed are also the exponential fit according to Eq. (\ref{eq:eq1}) and the modulation fit according to Eq.~(\ref{eq:eq2}). 
The inset shows the $\chi^2$ values  vs. the angular frequency $\omega$, for a fixed total decay constant $\lambda$  
and a variation of amplitude $a$  and phase $\phi$ .
}%
\label{fig2}%
\end{center}
\end{figure}

Figure~\ref{fig2} shows for altogether 3594 EC decays of H-like $^{142}$Pm$^{60+}$ ions the distribution of the decays per 0.96~seconds, 
as recorded by the 245~MHz resonator. 
The corresponding spectra were taken with a sampling time $\delta{t} = 32$~ms during 54~s long total observation periods
with 1688 independent measurements. 
The data analysis started 6.4~s after injection, which is the time needed for cooling the ions sufficiently, 
so that parent and daughter ions could unambiguously be distinguished. 
Either a strictly exponential distribution {model $M_0$, Eq.~(\ref{eq:eq1})}, was fitted to the data, 
or a periodically modulated exponential {model $M_1$, Eq.~(\ref{eq:eq2})}.
Both fits are displayed in Fig.~\ref{fig2}.

The modulation fit according to Eq.~(\ref{eq:eq2}) yields a total decay constant $\lambda_R = 0.0130(8)$~s$^{-1}$, 
an angular frequency $\omega_R = 0.884(14)$~s$^{-1}$ (period $T_R = 7.11(11)$~s), 
an amplitude $a_R = 0.107(24)$, and a phase $\phi_R = + 2.35(48)$~rad. 
The decay constant $\lambda$, the angular frequency $\omega$ and the period $T$ are related to the laboratory system. 
All elements of the correlation matrix are smaller than 0.1, except for the correlation coefficient between 
phase $\phi$ and angular frequency $\omega$ which amounts to $-0.85$. 
The inset shows the $\chi^2$ values vs. the angular frequency $\omega$, 
for a fixed total decay constant $\lambda$ and a variation of amplitude $a$ and phase $\phi$.

%
%

\begin{figure}[t!]
\begin{center}
\centering\includegraphics[angle=-0,width=0.45\textwidth]{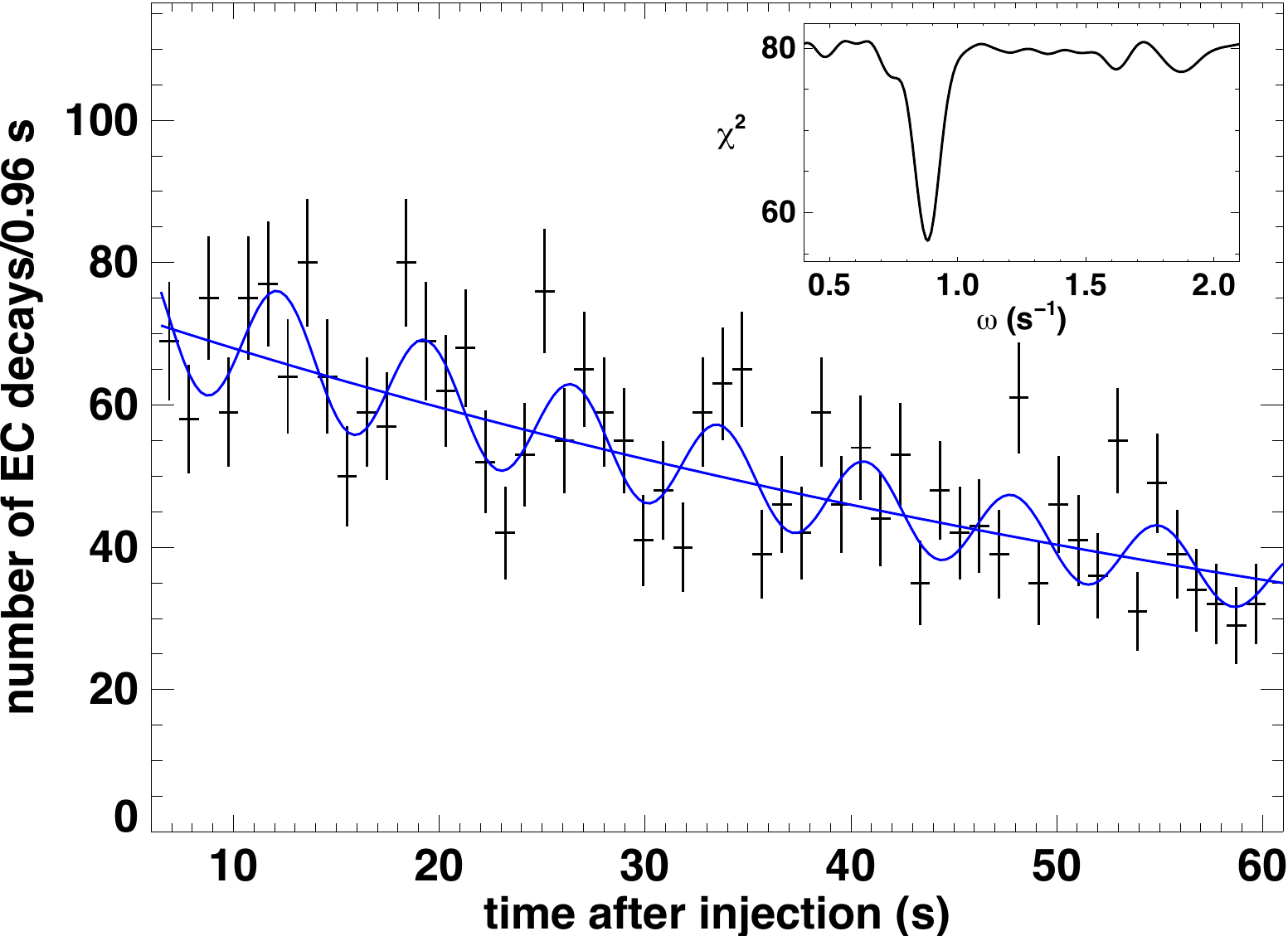}
\caption{
Number of EC decays per 0.96~s of H-like $^{142}$Pm$^{60+}$ ions, recorded by the capacitive pick-up, 
vs. the time after injection of the ions into the storage ring ESR. 
Displayed are also the exponential fit according to Eq. (\ref{eq:eq1}) and the modulation fit according to Eq.~(\ref{eq:eq2}).
The inset shows the $\chi^2$ values vs. the angular frequency $\omega$, 
for a fixed total decay constant $\lambda$ and a variation of amplitude  $a$  and phase $\phi$.
}%
\label{fig3}%
\end{center}
\end{figure}

Figure~\ref{fig3} shows for altogether 3098 EC decays the number of decays per 0.96~seconds vs. the time after injection, 
as recorded by the capacitive pick-up detector within the same {\it consecutive} measuring cycles as analyzed for the 245~MHz resonator. 
For this detector, however, only injections yielding at most two EC decays could be used for a safe analysis.
The corresponding spectra were taken with a sampling time $\delta{t} = 64$~ms during 54~s long total observation periods
with 844 independent measurements.
Again both fits are displayed in Fig.~\ref{fig3}.
%
%

The modulation fit according to Eq.~(\ref{eq:eq2}) (see Fig.~\ref{fig3}) yields a total decay constant $\lambda_P = 0.0126(7)$~s$^{-1}$, 
an angular frequency $\omega_P = 0.882(14)$~s$^{-1}$ (period $T_P = 7.12(11)$~s), an amplitude $a_P = 0.134(27)$, 
and a phase $\phi_P = + 1.78(44)$~rad. 
For this detector only the {\it appearance time} of the cooled daughter ion could be observed. 
The inset shows the $\chi^2$ values vs. the angular frequency $\omega$, 
for a fixed total decay constant $\lambda$ and a variation of amplitude $a$ and phase $\phi$.

A separate analysis has been performed of those EC decays which were detected {\it simultaneously} by both detectors. 
The smaller number of 3098~EC decays recorded by the capacitive pick-up with respect to the 3594~EC decays found 
with the 245~MHz resonator is almost entirely caused by the fact that in the case of the capacitive pick-up only 
files with at most two EC~decays could be safely analyzed, as mentioned above. 
For this comparison, still about 150 decays out of the 3098~EC decays have to be omitted, 
by ignoring those injections where the 245~MHz resonator was not triggered  (e.g., during the transfer of data) 
and, in particular, by omitting the data of all injections where not exactly the same number of EC decays was identified in both detectors. 
Hence, in the final sample both detectors see the random process of an EC decay fully correlated, for each single event. 
However, the detection of these events takes place quite differently: whereas the true decay time is revealed 
within 32~ms by the 245~MHz resonator (see Fig.~\ref{spectrum3d}), with the capacitive pick-up only 
the ``appearance'' of the already cooled daughter nucleus can be safely registered. 
Thus, for each EC decay recorded ``simultaneously'' by both detectors a time difference arises 
which corresponds to the cooling time of the recoiling daughter nucleus. 
The latter depends on the emission angle of the neutrino with respect to the beam direction, 
as visualized in Fig.~\ref{spectrum3d}. 
%
\begin{figure}[t!]
\begin{center}
\centering\includegraphics[angle=-0,width=0.45\textwidth]{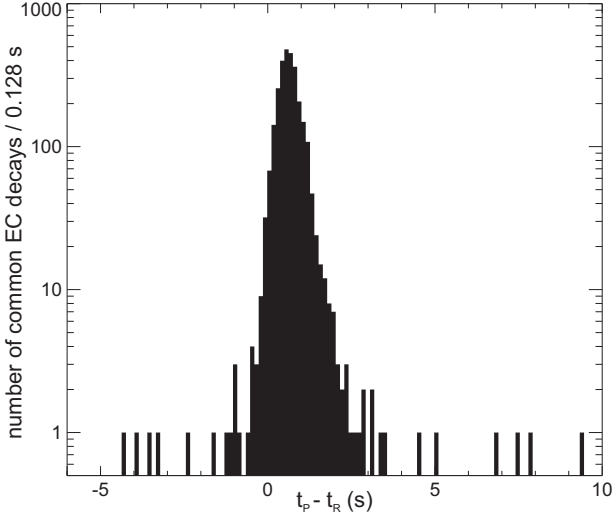}
\caption{
Distribution of the differences of times at which common EC decays have been recorded 
by the capacitive pick-up ($t_P$) and the 245~MHz resonator ($t_R$), respectively. 
The time binning is 128~ms.
}%
\label{newfig4}%
\end{center}
\end{figure}
%
The distribution of the recorded time differences of each of the correlated EC decays is shown in Fig.~\ref{newfig4}.

About 99 per cent of the central points are appearing within $-0.46~{\rm s} < \Delta t = t_P - t_R < + 2.70~{\rm s}$, 
with a mean time difference $\langle \Delta t\rangle = 669$~ms and a FWHM of 590~ms. 
The observed distribution convincingly demonstrates the strict correlation of almost all EC events, 
recorded simultaneously but independently by the two detectors. 
For 29 EC-events is $\Delta t < 0$, and for 14 EC decays one observes $\Delta t > + 2.70$~s 
(in total a  fraction of $1.5 \cdot 10^{-2}$), pointing to a small background of uncorrelated events 
which is caused, most probably, by systematic errors in the analysis of the pick-up data due to the poor signal-to-noise characteristics of this detector. 

For this selection of simultaneously recorded EC decays, independent fits of the events observed 
by the 245~MHz resonator and the capacitive pick-up, respectively, have been performed. 
They yielded an angular frequency $\omega_R = 0.887(11)$~s$^{-1}$, an amplitude $a_R = 0.147(28)$, 
and a phase $\phi_R = 2.55(37)$ radian for the 245~MHz resonator, 
and $\omega_P = 0.879(11)$~s$^{-1}$, $a_P = 0.161(28)$, $\phi_P = 1.98(35)$ radian for the capacitive pick-up. 
The phase difference $\Delta \phi = \phi_R - \phi_P$ amounts to $\Delta \phi = 0.57(50)$ radian 
which corresponds to a mean time delay of $\langle \Delta t \rangle = 646$~ms and reproduces, thus,  
rather exactly the mean time difference $\langle \Delta t \rangle = 669$~ms as observed for 
the 99 percentile of the distribution of the time differences as shown in Fig.~\ref{newfig4}.
The fit values of both, the complete and the correlated data sets of the two detectors, are further discussed in 
Section~\ref{S:synopsis} and summarized in Table~\ref{tablex}.

In order to assess the significance of these results the evidence criteria are used in favour of a certain model.
We compare the reliability 
of the two models $M_0$ (strictly exponential distribution) and $M_1$ (periodically modulated exponential distribution) 
for describing the measured time distribution of the EC decays. 
For this purpose we exploited the ``Akaike Information Criterion ($AIC$)''~\cite{r21}. 
For $\chi^2$ minimization, the $AIC$ value of a model $M_i$, 
corrected for a number $k_i$ of fit parameters and a number $n_i$ of data points, is defined as~\cite{r21,r22}:
\begin{equation}
AIC (M_i) = 2k_i +\chi^2_i + 2k_i \frac{k_i + 1}{n_i - k_i-1} - 2C.
\label{eq:eq3}
\end{equation}
Here $C$ is a constant, if the number $n_i$ of data points is the same for all compared models $M_i$. 
The $AIC$-difference $\Delta_{AIC}$ of two models $M_i$ and $M_{\rm min}$:
\begin{equation}
\Delta_{AIC}  =  AIC (M_i) - AIC (M_{\rm min})
\end{equation}
estimates the information loss experienced if using model $M_i$ instead of the model with the smallest $AIC$ value, $M_{\rm min}$. 
In our case model $M_1$ (modulated exponential distribution) is the model with the smaller $AIC$ value. 
For the EC decays of the complete data set, one obtains $\Delta_{AIC}$  from the corresponding  $\chi^2_i$ values, 
the number $k_i$ of parameters and from the number $n_i$ of data points (see Eq.~(\ref{eq:eq3})) as: 
\begin{eqnarray}
\label{eqr}
\Delta_{AIC} = AIC(M_0) - AIC(M_1) = 12.46;\\
\label{erp}
\Delta_{AIC} = AIC(M_0) - AIC(M_1) = 17.24,                                                                         
\end{eqnarray}
where Eqs.~(\ref{eqr}) and (\ref{erp}) stand for the case of the 245~MHz resonator and the capacitive pick-up, respectively.
With these numbers the Akaike weights $w(M_i)$ are calculated according to~\cite{r22}, 
which are the ``weights of evidence'' of a specific model among all models considered:                          
\begin{equation}
\label{eqr1}
w(M_0) = \frac{e^{{-\Delta_{AIC}}/{2}}} {1 +  e^{-\Delta_{AIC}/2}}~~~{\rm and}~~~ 
w(M_1) = \frac{1}{1 +  e^{{-\Delta_{AIC}}/{2}}}.
\end{equation}
Hence for the 245~MHz resonator we obtain for the complete data set:
\begin{equation}
w(M_0) =  1.97 \cdot 10^{-3}~~~{\rm and}~~~w(M_1)= 0.998,                                         
\end{equation}
and for the capacitive pick-up:
\begin{equation}
w(M_0)  =  1.80 \cdot 10^{-4}~~~{\rm and}~~~w(M_1) = 0.9998.                                         
\end{equation}
The corresponding values for the correlated data set 
are $w(M_0)=2.20\cdot10^{-5}$, $w(M_1)=0.99998$ for the 245~MHz resonator and
$w(M_0)=2.55\cdot10^{-5}$, $w(M_1)=0.99997$ for the capacitive pick-up.

In addition, a simulation based on the {\it fit values of a strictly exponential distribution} was performed 
in $10^5$ Monte Carlo trials for the complete data set, 
where the weights of evidence $w(M_0)$ and $w(M_1)$ have been calculated for each Monte Carlo experiment. 
One obtains for the 245~MHZ resonator, if the underlying model is $M_0$, 
a probability $r < 4.3 \cdot 10^{-3}$ to get just by chance a value $\Delta_{AIC}  = AIC (M_0) - AIC (M_1) \ge 12.46$ (see Eq.~(\ref{eqr})). 
For the capacitive pick-up with  $\Delta_{AIC} = 17.24$ (see Eq.~(\ref{erp})) the corresponding probability amounts to $r < 1.1\cdot 10^{-3}$.

For both detectors we randomly selected from the complete data sets subsets of different fractions of the total number of recorded EC decays. 
Each subset, down to a fraction of 0.5, was constructed by 50 trials. 
We found that in both cases the angular frequency as well as the amplitude stay constant within their (increased) error margins, 
whereas the $\Delta_{AIC}$ values, relevant for the weight of evidence and thereupon for the model selection, 
decrease smoothly for the 245~MHz resonator from $\Delta_{AIC} = 12.46$ (Eq.~\ref{eqr}) 
for the complete data set to $\Delta_{AIC} \approx 5$ for a fraction of 0.5. 
For the capacitive pick-up the corresponding values decrease from $\Delta_{AIC} = 17.24$ (Eq.~\ref{erp}) to $\Delta_{AIC} \approx 8$. 
These results prove, on the one hand, that the modulation will be found in any (not too small) 
partition but, on the other hand, that the weights of evidence become smaller in proportion to the decreasing decay statistics, as expected.

\subsection{Results for the $\beta^+$ decay of hydrogen-like $^{142}$Pm$^{60+}$ ions}
\label{s42}
Also the $\beta^+$ decay branch has been inspected with the 245~MHz resonator, 
because it is of utmost importance to verify 
whether in the three-body $\beta^+$ decay also a 7~s modulation occurs 
comparable to the one observed in the two-body EC decay. 
Besides via $\beta^+$ decay, the parent ion can also disappear by capturing or losing an electron in the cooler or in the residual gas. 
This loss constant is small and amounts in the present experiment to  $\lambda_{\rm loss} = 5(1)\cdot10^{-4}$~s$^{-1}$ 
which is more than one order of magnitude smaller than the corresponding decay constants of  both the $\beta^+$ and the EC decay. 
Nevertheless, since a sequence of atomic-charge exchange reactions and $\beta^+$-decays can feed the same daughter ions $^{142}$Nd$^{60+}$,
the orbits of the corresponding electron pick-up or electron-loss reaction products were blocked in the ESR by mechanical scrapers.
The latter is easily achieved since the change of the atomic charge state results in a large alteration of the revolution frequency~\cite{LiBo}.
In this case, however, also the $\beta^+$ decay daughter ions $^{142}$Nd$^{59+}$ are removed from the ESR.
Therefore, we could only observe the {\it disappearance} of a parent ion {\it without} detecting the 
simultaneous appearance of its H-like $^{142}$Nd$^{59+}$ daughter ion. 
Furthermore, because the identification of $\beta^+$ decays by means of the capacitive pick-up is quite difficult, 
the present analysis was restricted to $\beta^+$ decay-data recorded by the 245~MHz resonator. 
The corresponding spectra were taken with a sampling time $\delta t = 32$~ms during a 50~s long total observation period. 

For the purpose of a reliable comparison of $\beta^+$- and EC-decay data, the same observation cycles have been 
analyzed with respect to $\beta^+$ decays, in which the 3594 consecutive EC decays occurred. 
But only cycles with at {\it most four} injected parent ions have been used, where the loss of one ion 
could safely be identified, at the expense of a reduced statistics. 
Unambiguous detection of the disappearance of a parent ion is only 
guaranteed after completed cooling of {\it all} parent ions. 
Since this takes -- in a few cases -- up to 10~seconds, the analysis of the $\beta^+$ decays 
had to be restricted, hence, to times $t \ge 10$~s. 

%
\begin{figure}[t!]
\begin{center}
\centering\includegraphics[angle=-0,width=0.45\textwidth]{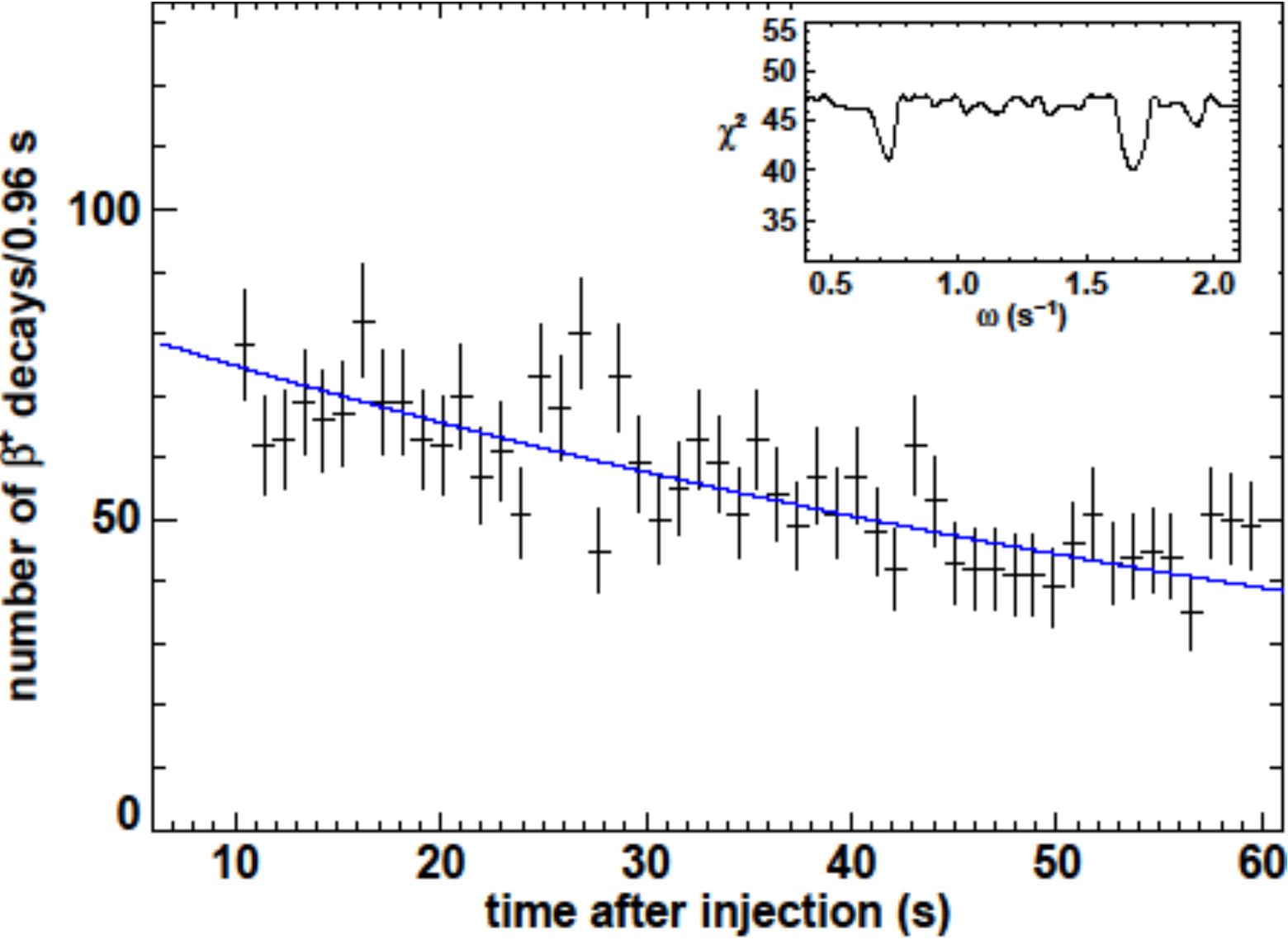}
\caption{
Number of $\beta^+$ decays per 0.96~s of hydrogen-like $^{142}$Pm$^{60+}$ ions, recorded by the 245~MHz resonator, vs. the time after 
injection of the ions into the storage ring ESR. 
Also shown is the exponential fit according to Eq.~(\ref{eq:eq1}). 
The inset displays, for a supposed modulation, the $\chi^2$  
values vs. the angular frequency $\omega$, for a fixed total decay constant $\lambda$ and a  variation of amplitude $a$ and phase $\phi$.
}%
\label{fig4}%
\end{center}
\end{figure}
%

Figure~\ref{fig4} shows the $\beta^+$ decay rate per 0.96~seconds for altogether 2907 $\beta^+$ decays 
as a function of the time after injection. 
Besides a strictly exponential fit (model $M_0$) yielding a total decay constant $\lambda = 0.0130(8)$~s$^{-1}$ (see Fig.~\ref{fig4}), 
also a modulation fit has been performed. 
For a presumed modulation of the $\beta^+$ decays, the parameter $\lambda_{\rm EC}$ in Eq.~(\ref{eq:eq1}) 
has to be replaced by $\lambda_{\beta^+}$ and, correspondingly, $\lambda_{\rm EC} (t)$ in Eq.~(\ref{eq:eq2})  by $\lambda_{\beta^+} (t)$. 
The inset of Fig.~\ref{fig4} shows, for a supposed modulation with fixed 
total decay constant  $\lambda$, the $\chi^2$ values vs. $\omega$ for a variation of amplitude $a$ and phase $\phi$. 
We analyzed also whether the three-body $\beta^+$ decay exhibits a 
significant modulation at an angular frequency near to the one found for the EC decays. 
For this purpose, the angular frequency region $0.81~{\rm s}^{-1} \le \omega \le 0.95~{\rm  s}^{-1}$, 
the range of $\pm$~five standard deviations of the angular frequency observed for the 
EC decay with the 245~MHz resonator $(\omega = 0.884(14)~{\rm s}^{-1})$, has been scrutinized in detail.

The fit yields as the $\chi^2$ minimum in this range of angular frequencies 
the value $\omega = 0.907~{\rm s}^{-1}$, a phase $\phi = 3.1(25)$ and a modulation amplitude $a = 0.027(27)$, compatible with $a = 0$. 
This corroborates with somewhat improved decay statistics the preliminary result $a = 0.03(3)$ reported in Refs.~\cite{PK-NPA09, PK-PPNP10}. 

The $AIC$ has also been exploited for the $\beta^+$ data. 
For the fit values given above we find as  $\Delta_{AIC}$-value: 
\begin{equation}
\Delta_{AIC}  =  AIC (M_1) - AIC (M_0) =5.74,
\end{equation}
since in this case $M_0$ is the model $M_{\rm min}$ with the smallest information loss. 
Likewise, for the weights of evidence one gets:
\begin{equation}
w(M_1) =0.054~~~{\rm and}~~~ 
w(M_0) =0.946.
\end{equation}

\subsection{Synopsis of the results and conclusions }
\label{S:synopsis}

By means of a newly available 245~MHz resonator the detection sensitivity of circulating, cooled, 
highly-charged ions was improved by about two orders of magnitude compared with the previously used conventional capacitive pick-up. 
This enabled a precise determination of the true EC decay time of H-like $^{142}$Pm$^{60+}$ ions, 
and an unprecedented time- and frequency-resolved measurement of the 
kinematics of the recoils (and, hence, of the neutrinos) just from  the moment of their generation. 
Therewith, the true decay time could be determined unambiguously for the two-body 
EC decay with an accuracy of approximately 32~ms. 

The whole of recorded data {\it does not} show modulations which might be caused by failures of the injection hardware as outlined in Section~\ref{s:s3}.
A series of 7125 consecutive injections could be found for which no hardware malfunctions were documented.
This series has been used for the data analysis (``complete data sets'').
Therewith the EC decay-statistics could be augmented with respect to the previous experiment but, 
owing to the technical problems, not as much as originally intended. 
For the same reason, also the statistics of the three-body $\beta^+$ decays could only moderately be improved. 
Furthermore, a subset of about 2940 EC decays has been analyzed, where each decay was ``simultaneously'' registered by both, 
the 245~MHz resonator and the capacitive pick-up detector (``correlated data sets'').
The fit results, the $\Delta_{AIC}$ values and the ``weights of evidence'' $w(M_i)$ 
are summarized for both detectors and for both data sets in Table~\ref{tablex}.
For all four EC spectra the weights of evidence $w(M_1)$ are larger than $0.998$. 

\begin{table*}[t!]
\caption{%
The table shows for both detectors the $\chi^2/{\rm DoF}$-values and fit parameters 
for angular frequency $\omega$, amplitude $a$ and phase $\phi$ of a supposed modulation on top of an exponential decay. 
In the third and fourth lines these values are given for the EC decays in the complete data sets (3594~EC decays for the 245~MHz resonator, 
3098~EC decays for the capacitive pick-up), and in the seventh and eight lines for the data sets of the simultaneously observed EC decays. 
The last line provides the corresponding values for the $\beta^+$ decays recorded by the 245~MHz resonator (see Section~\ref{s42}). 
The last three columns display the observed $\Delta_{AIC}$-values and the ``weights of evidence'' $w(M_i)$, 
which are important as criteria for model selection (see Sections~\ref{s41} and \ref{s42}).  
In our case we compare a model $M_0$ of a strictly exponential distribution of the decays 
with a model $M_1$ of a periodically modulated exponential distribution.
\label{tablex}}%
\begin{center}
\begin{tabular}{cccccccc}
EC decay    & $\chi^2/{\rm DoF}$& $\omega$ (s$^{-1}$) & $a$ & $\phi$ (rad.) & $\Delta_{AIC}$ ($M_{\rm min}=M_1$) & $w(M_0)$ & $w(M_1)$ \\ 
complete data sets & & & & & & & \\
\hline
245~MHz        & 33.6/52 & 0.884(14) & 0.107(24) & 2.35(48) & 12.46 & $1.97\cdot10^{-3}$ & 0.998\\
capac. pickup & 56.6/51& 0.882(14) & 0.134(27) & 1.78(44) & 17.24 & $1.80\cdot10^{-4}$ & $0.9998$\\
\\
EC decay     \\ 
correlated data sets & & & & & & & \\
\hline
245~MHz        & 33.9/51 & 0.887(11) & 0.147(28) & 2.55(37) & 21.45 & $2.20\cdot10^{-5}$ & 0.99998\\
capac. pickup & 65.5/51& 0.879(11) & 0.161(28) & 1.98(35) & 21.15 & $2.55\cdot10^{-5}$ & $0.99997$\\
\\
$\beta^+$ decay&    &  &  &  &  $\Delta_{AIC}$ ($M_{\rm min}=M_0$)& & \\ 
\hline
245~MHz        & $46.5/48^{\ddag}$& $0.907(20)^*$ & $0.027(27)^*$ & $3.1(2.5)^*$ & 5.74 & 0.946 & $5.36\cdot10^{-2}$\\
\hline
\end{tabular}\end{center}
$^\ddag$ - fit of a purely exponential decay \\
$^*$ - fit values with the constraint $0.81~{\rm s}^{-1} \le \omega \le 0.95~{\rm s}^{-1}$  
\end{table*}

%
Based on  these weights, one can conclude that the probability of model $M_1$ to be the best model 
(in the sense of the Kullback-Leibler divergence) in the model set $\{M_0, M_1\}$ is greater than 99.8 \%. 
This conclusion is strongly endorsed by the result of Monte Carlo 
simulations, which show for both detectors 
a probability $r < 4.3\cdot10^{-3}$ that a value $\Delta_{AIC}  = AIC(M_0) - AIC(M_1) \ge 12.46$ 
could occur {\it just by chance}, if the underlying model was $M_0$.

The four angular frequencies, observed by the two detectors in the complete as well as in the correlated data sets , 
are in perfect agreement with each other within their error margins (see Table~\ref{tablex}) 
and also with the published value $\omega = 0.885(31)$~s$^{-1}$ \cite{Osc-PLB}. 
In the complete data sets the modulation amplitude obtained for the capacitive pick-up ($a_P = 0.134(27)$) 
is  by one standard deviation larger than the amplitude extracted from the data of the 245~MHz resonator ($a_R = 0.107(24)$). 
For the correlated data sets,  both amplitudes become considerably larger ($a_P({\rm correlated~set}) = 0.161(28)$, 
$a_R({\rm correlated~set}) = 0.147(28)$) with respect to the complete data sets, 
whereas their difference is lowered to 0.5 standard deviations.  
For the 245~MHz resonator we notice that in the correlated data set, 
with at most two registered EC decays per injection, 
the amplitude is increased by more than 1.5 standard deviations with respect to the complete data set, 
where up to six EC decays per injection have bee observed. 

This raises the question whether the modulation parameters could be influenced by the number of injected ions. 
It is well-established that a few stored and cooled ions build up a phase transition to a coupled and ordered system,  
(``linear Coulomb crystals'' \cite{chains}), where any intra-beam scattering disappears. 
Whether the coupling of unstable, stored and cooled  ions to the (ion number dependent) 
eigenmodes of this system could finally influence even their weak two-body decay, 
is a question worthy of examination which has been hitherto neither theoretically nor experimentally addressed.  
Concerning our experiments, the lack of sufficient decay statistics as function of the number of stored parent ions  
prevents any feasible proof of this speculative thought.

In the correlated set the amplitude $a_P = 0.161(28)$ found 
with the pick-up detector is indeed still larger than the amplitude $a_R = 0.147(28)$ observed with 
the 245~MHz resonator, though with a considerably smaller 
difference than it was found for the complete data sets ($a_R({\rm complete~set}) = 0.107(24)$, $a_P({\rm complete~set}) = 0.134(27)$). 
Now, the ``appearance'' times recorded by the pick-up originate from a convolution 
of the true decay times with the distribution of the cooling times. 
Thus, a larger amplitude has to be expected for the 245~MHz resonator which provides the true decay times (see Fig.~\ref{spectrum3d}). 
However, the ``inverted'' outcome, showing still a slightly smaller amplitude for the 245~MHz resonator is, 
within statistical uncertainty,  not in contradiction to this well-considered anticipation, 
when taking into account the small width of the cooling time distribution with respect to the modulation period of 7~s and 
keeping in mind 
the small difference of the amplitudes of only 0.5 standard deviations.
%
We note, moreover, that -- in spite of the smaller amplitude -- the weight of evidence $w(M_1)$ 
is slightly larger for the correlated data set taken by the 245~MHz resonator than the corresponding weight for the correlated data
recorded by the capacitive pick-up (see Table~\ref{tablex}).

All the amplitudes of our four data sets differ, however, distinctly from the published amplitude $a = 0.23(4)$ \cite{Osc-PLB}. 
It has to be emphasized that the latter is related to a total observation time $t_{\rm obs} = 33$~s, 
whereas all present results refer to $t_{\rm obs} = 60$~s. 
But a scan of the new data as a function of $t_{\rm obs}$ could not provide 
a statistically significant dependence of the amplitudes on the observation time. 
A noteworthy point -- in particular in connection with the reasoning above -- is the difference of the average number
of injected ions, which was near to four in the present experiment while previously it was only about two.
Yet for the present, the discrepancy of the amplitudes remains an unresolved puzzle.
 
Concerning the $\beta^+$ decay data (see Fig.~\ref{fig4}), recorded by the 245~MHz resonator, a maximum amplitude  $a = 0.027(27)$, 
arises within the investigated range of angular frequencies $0.81~{\rm  s}^{-1} \le \omega \le 0.95~{\rm  s}^{-1}$ at $\omega = 0.907~{\rm s}^{-1}$, 
compatible with $a = 0$ and in accordance with the previously reported value a = 0.03(3)~\cite{PK-NPA09, PK-PPNP10}.
Now one can probe whether or not this amplitude reflects the same 
underlying time distribution of the decays as does the amplitude $a = 0.107(24)$, derived for the EC decays. 
First of all, one might suppose that both, the EC- and the $\beta^+$- decay modes, do follow the same time distribution. 
Thereupon both amplitudes would also belong to a common, 
normally distributed set and their weighted mean $\langle a \rangle = 0.0715(180)$ 
could be assumed as a reliable first estimate of the true amplitude which also minimizes the $\chi^2$ value. 
Under this hypothesis a $\chi^2$ test on the two measured amplitudes has been performed. 
The resulting $\chi^2 = 4.90$ corresponds for one degree of freedom to a $p$-value of:
 \begin{equation}
\label{eq8}
 p\le 0.026
 \end{equation}
 
 Hence, one would observe such a difference (or a larger one) of the amplitudes only in 3\% 
 of the cases at the most, if the EC- and the $\beta^+$ amplitudes would originate from the same underlying time distribution of the decays. 
 Thus one can exclude, based on the $p$ value of Eq.~(\ref{eq8}), at a confidence level $> 0.97$  
 that this amplitude $a = 0.027(27)$ and the amplitude $a = 0.107(24)$, 
 as observed for the EC decay, belong to the same time distribution of the decays. 
 Furthermore, from the weights of evidence shown in Table~\ref{tablex} the ratio $R = w(M_1)/w(M_0) = 0.057$
 is derived, which yields, for the investigated range of angular frequencies, 
 the relative probability of a periodic modulation of the $\beta^+$ decay with respect to a strictly exponential distribution. 
 
 \section{Tentative interpretation of the results and outlook }

 Our results indicate, as the most important finding of the present experiment, that  a {\it significant 
 7-second modulation} of the weak decay of H-like $^{142}$Pm$^{60+}$ ions,  
 although existing  in connection with their {\it two-body} EC decays,  
 is {\it absent} in their   three-body  $\beta^+$ decays. 
 Thus, the data could point to a weak-interaction origin of the modulation which 
 is observed in connection with the monoenergetic electron-neutrinos from 
 EC decays, but not visible for the continuous neutrino spectrum of the $\beta^+$ decay branch (see also Ref.~\cite{Ivanov-PRL101}). 
 Finally, our findings also suggest that interferences occur in the two-body EC decay although the neutrinos are {\it not} directly observed.

 
The flavour eigenstate of the electron neutrino is known to be a superposition of mass eigenstates $m_j$ 
of massive neutrinos with slightly different energies \cite{Giunti, Ivanov-PRL103}. 
This could lead, hypothetically, to a time dependent interference pattern for its squared wave function, 
but only in case of the two-body EC decay with the creation of an (almost) monoenergetic electron neutrino.
 From energy and momentum conservation in every EC decay channel $p \to d + \nu_j$ 
 the energy of massive neutrino $\nu_j$ is given in the center of mass system by $E_j = (M_p^2 -M_d^2 + m_j^2) c^2/2M_p$ 
 which leads to an energy difference of the neutrino mass-eigenstates and a modulation frequency 
 $\omega_{ij} =  (E_i -E_j)2\pi/h = \Delta m_{ij}^2c^22\pi/2M_ph$, where $\Delta m_{ij}^2 = m_i^2 - m_j^2$ 
 is the squared mass difference of $\nu_i$ and $\nu_j$. 
 Note, that $\omega_{ij}$ determines also the recoil 
 energy difference of the observed daughter ions, 
 and is predicted to be proportional to the inverse of the parent mass $M_p$~\cite{PK-arxiv1311}. 

If the modulation, found in the EC decay, 
would be due to an interference of mass eigenstates, we deduce, in the approximation of two states,
 from the observed average value of the angular frequency of the complete data set, $\langle \omega_{12} \rangle = 0.883(10)$~s$^{-1}$, 
 a squared mass difference $\Delta m_{12}^2 = 2.19(3) \cdot 10^{-4}$~eV$^2$/c$^4$. 
 This value is nearly three times larger than $\Delta m_{12}^2 = 7.59(21) \cdot 10^{-5}$~eV$^2$/c$^4$ 
 determined by the KamLAND collaboration for antineutrinos emitted by fission products~\cite{Abe}. 
 
It has often been stated~\cite{Giunti, Kienert08, Merle10, Cohen} that in the case that
``the neutrinos are not observed'' the {\it squares} of the individual transition amplitudes must be added and that, 
therefore, the modulations observed here cannot be caused by neutrino properties. 
Indeed neutrinos can, in principle, not be observed between the time of the production of the 
parent  and the observation of the daughter ion. 
It was pointed out that for this reason one has to sum the transition amplitudes over {\it all} neutrino 
flavours~\cite{Gal, Yazaki, PK-ejtp09}. 
Then the interference indeed disappears for a unitarian flavor mixing matrix. 
This property could be missing if, for instance, ``sterile'' neutrinos and/or CP-violating phases do exist~\cite{Gal, Yazaki, PK-ejtp09}.

If our considerations are correct, two-body weak decay experiments with
highly ionized ions by means of single-ion decay spectroscopy 
would be a new method to study exciting basic properties of massive neutrinos and antineutrinos. 
Of special interest for fundamental tests would be a direct comparison 
of anticipated modulation parameters for the EC decay, 
generating monoenergetic electron neutrinos, and its time-mirrored process, the bound-state $\beta$ decay, 
where monoenergetic {\it antineutrinos} are created. 
Both decay modes can occur in the two-body $\beta$ decays of some odd-odd nuclei in  
H-like or completely ionized atoms, such as $^{108}_{~47}$Ag or $^{112}_{~49}$In.

\section*{$^\dag$ In Memoriam}

We mourn for Paul Kienle, the former Scientific Director of GSI in Darmstadt and Professor emeritus of the 
TUM (Technische Universit{\"a}t M{\"u}nchen). He passed away on January 29, 2013. 
Besides many other outstanding achievements we owe him the construction and successful operation 
of both, the fragment separator FRS and the storage ring ESR, for more than 20 years now. 
His tireless passion and absolute devotion to physics stays as a hardly 
accessible example for all of us. 
Until the very last days of his life Paul Kienle worked on the finalization of this manuscript
and we tried to preserve his latest suggestions to this paper. 
His death is an irreplaceable loss for our collaboration and for the physics community.

\section*{Acknowledgements}

We would like to thank the GSI accelerator and FRS teams 
for the excellent technical support and their steady help. 
In particular we are grateful to P.~Petri for his decisive contribution to the perfect operation of the new 245~MHz resonator. 
We thank Profs. D.~Liesen, K.~Langanke and H.~St{\"o}cker for kindly providing additional beam time.  
We are indebted to Profs. K.~Blaum, P.~Braun-Munzinger, H.-G.~Essel, M.~Faber, H.~Feldmeier, A.~Gal, 
F.~Giacosa, W.~F.~Henning, A.~N.~Ivanov, C.~Kozhuharov, H.~Lipkin, H.~Kleinert, H.-J.~Kluge, R.~Kr{\"u}cken, G.~M{\"u}nzenberg,
Z.~Patyk, C.~Peschke, C.~Rappold, H.~Rauch, G.~Rempe, H.~Schmidt, D.~Schwalm, V. Soergel, B. Stech and K.~Yazaki for many fruitful discussions. 
This research was partially supported by the DFG cluster of 
excellence ``Origin and Structure of the Universe'' of the Technische Universit{\"a}t M{\"u}nchen,
by the BMBF grant in the framework of the
Internationale Zusammenarbeit in Bildung und Forschung Projekt-Nr.
01DO12012, by the Helmholtz-CAS Joint Research Group HCJRG-108, and
by the External Cooperation Program of the Chinese Academy Sciences Grant No. GJHZ1305.
I.~D. is funded by the Helmholtz Association via the Young Investigators Project VH-NG 627.
T.~Y., T.~S., F.~S., T.~I. acknowledge the support by the Japanese Ministry of Education, Science,
Sport and Culture by Grant-In-Aid for Science Research under Program No. A 19204023.
C.~B. acknowledges support by BMBF (contracts 06GI911I and 06GI7127/05P12R6FAN) 
and by the Alliance Program of the Helmholtz Association (HA216/EMMI).
M.~S.~S. acknowledges HIC-for-FAIR through HGS-HIRE.

\end{document}